\documentclass[12pt]{article}

\usepackage{amsmath,amsfonts,amssymb}

\def\mH{\mathcal{H}}

\def\bx{\mathbf{x}}
\def\by{\mathbf{y}}
\def\bz{\mathbf{z}}

\newcommand{\hg}{\hat{g}}

\newcommand{\mG}{\mathcal{G}}

\newcommand{\bT}{\mathbf{T}}

\def\pb #1{\left\{#1\right\}}

\begin{document}

\begin{titlepage}

\rightline{\footnotesize{}} \vspace{-0.2cm}

\begin{center}

\vskip 0.4 cm

\begin{center}
{\Large{ \bf Hamiltonian Analysis of
 Lagrange Multiplier Modified Gravity}}
\end{center}

\vskip 1cm

{\large Josef Kluso\v{n}$^{}$\footnote{E-mail:
{\tt klu@physics.muni.cz}}
}

\vskip 0.8cm

{\it
Department of
Theoretical Physics and Astrophysics\\
Faculty of Science, Masaryk University\\
Kotl\'{a}\v{r}sk\'{a} 2, 611 37, Brno\\
Czech Republic\\
[10mm]
}

\vskip 0.8cm

\end{center}

\begin{abstract}
We develop Hamiltonian formalism for Lagrange
Multiplier Modified Gravity. We further
calculate  the
Poisson brackets between constraints and
we show that they coincide with the algebra
of constraints in Hamiltonian formulation of
 General Relativity.

\end{abstract}

\bigskip

\end{titlepage}

\newpage

\section{Introduction}\label{first}
One of the most important problem in present cosmology is
the  understanding of the origin of the late-time cosmic acceleration
(the so-called Dark Energy (DE) epoch). Recently new
interesting model DE model was proposed in
\cite{Gao:2010gj,Lim:2010yk}.
This model  consists of two scalar fields
where one of scalars represents the Lagrange multiplier.
The multiplier puts constraint   on the second
scalar field and as a result the theory contains
singe degrees of freedom. It was shown that
the energy of the system flows along time-like geodetic
that is similar to the dust, however the theory
contains non-zero energy.  The behavior of this
system suggests that it can be
 natural candidate for unification of Dark Energy and
Dark Matter. The cosmological implications of
these models were then analyzed in
\cite{Du:2010rv,Feng:2010tf,Cai:2010zm}.
The role of Lagrange multipliers in the context of
$f(R)$ gravities was studied in
\cite{Capozziello:2010uv}.
 Moreover, the Lagrange multipliers in the context
 of modified gravity may improve the ultraviolet
 properties of the covariant Ho\v{r}ava-Lifshitz
 gravity  \cite{Nojiri:2009th}
 leading to its   renormalizability conjecture
\cite{Nojiri:2010kx,Nojiri:2010tv}.

As was shown in all these papers
the presence of the Lagrange multipliers in the
action has strong impact on the form of the
resulting equations of motions. Then it is
natural to ask the question how the presence
of Lagrange multipliers  modifies   Hamiltonian
structure of given theory. Moreover, we would
like to see whether the Hamiltonian of these
systems is again given as a   linear combination of
 constraints and whether these constraints are
the first class and their Poisson algebra
respects the basic principles of geometrodynamics
\cite{Isham:1984sb,Isham:1984rz,Hojman:1976vp}.
It turns out that
Hamiltonian structure  of given theory
is  very interesting.
 We show that
the presence of the
first scalar field that plays the role of the
 Lagrange multiplier
implies an existence of the second class
constraints.
Then   after their solving  we find
the Hamiltonian equations of motions
for the second scalar field  that are  autonomous
in the sense that the time evolution of
the scalar field does not depend on
its conjugate momenta.
 Such systems were
studied in the past especially in the context
of the 't Hooft deterministic approach to
quantum mechanics
\cite{'tHooft:2001ct,'tHooft:1999gk,'tHooft:1999bx,Blasone:2009yp}.
We also find that the resulting theory is a fully
constrained system with the algebra of constraints
that has the same form as in General Relativity.

As the second example of the Lagrange multiplier
modified theory we consider the gravity action
introduced in \cite{Capozziello:2010uv}. This
action is the Lagrange modification of $F(R)$
gravity theories
\footnote{For review, see
\cite{DeFelice:2010aj,Faraoni:2008mf,Nojiri:2006ri}.}.
We show that the resulting Hamiltonian is given
as a linear combination of constraints and
 has similar
structure as the Hamiltonian of  $F(R)$
gravities
\cite{Deruelle:2009pu,Chaichian:2010zn}
\footnote{For related works, see
\cite{Ezawa:1998ax,Ezawa:1999ty}.}.
However there is an important
difference that follows the fact that
the presence of the Lagrange multiplier
implies that the original auxiliary
fields become dynamical in Hamiltonian
formulation. We further determine the
Poisson brackets between
 constraints. We show that the algebra
of these constraints takes exactly the same form
as in \cite{Isham:1984sb,Isham:1984rz,Hojman:1976vp}.
In other words we explicitly prove the
consistency of Lagrange modified theories of
gravity from the Hamiltonian point of view.

Let us summarize our results. We study the
Lagrange multiplier modified theories with emphasis
on their Hamiltonian formalism. We find
that the resulting Hamiltonian
is again given as a linear combination of the
first class constraints. We show that the
Poisson brackets of these constraints have
the same form as in General Relativity.

This paper is organized as follows. In the next
section (\ref{second}) we perform the Hamiltonian
formulation of the General Relativity action
together with the
Lagrange multiplier modified scalar field action.
We find corresponding Hamiltonian and diffeomorphism
constraints and calculate their algebra. In section
(\ref{third}) we study the Lagrange multiplier
modified action introduced in
\cite{Capozziello:2010uv}. We again  determine corresponding
Hamiltonian. Then we calculate the Poisson
brackets of the secondary constraints  and we find
that they  take exactly the same form as in
General Relativity.

\section{Lagrange Multiplier Modified Scalar
Field Action}\label{second}
In this section we develop the Hamiltonian
formalism for Lagrange multiplier modified
scalar field action. We study the form of the
action that was introduced in
\cite{Capozziello:2010uv}
\begin{equation}
S=\int d^{(D+1)}x \sqrt{-\hg}[
{}^{(D+1)}R(\hg)-\frac{\omega(\phi)}{2}
\hg^{\mu\nu}\partial_\mu\phi
\partial_\nu\phi-V(\phi)-\lambda
[\frac{1}{2}\hg^{\mu\nu}
\partial_\mu\phi\partial_\nu\phi+U(\phi)]] \ .
\end{equation}
Let us explain our notation.
We  consider $D+1$ dimensional
manifold $\mathcal{M}$ with the
coordinates $x^\mu \ , \mu=0,\dots,D$
and where $x^\mu=(t,\bx) \ ,
\bx=(x^1,\dots,x^D)$. We presume that
this space-time is endowed with the
metric $\hat{g}_{\mu\nu}(x^\rho)$ with
signature $(-,+,\dots,+)$. Suppose that
$ \mathcal{M}$ can be foliated by a
family of space-like surfaces
$\Sigma_t$ defined by $t=x^0$. Let
$g_{ij}, i,j=1,\dots,D$ denotes the
metric on $\Sigma_t$ with inverse
$g^{ij}$ so that $g_{ij}g^{jk}=
\delta_i^k$.
 We  introduce  the
future-pointing unit normal vector
$n^\mu$ to the surface $\Sigma_t$. In
ADM variables we have
$n^0=\sqrt{-\hat{g}^{00}},
n^i=-\hat{g}^{0i}/\sqrt{-\hat{g}^{
00}}$. We also define  the lapse
function $N=1/\sqrt{-\hat{g}^{00}}$ and
the shift function
$N^i=-\hat{g}^{0i}/\hat{g}^{00}$. In
terms of these variables we write the
components of the metric
$\hat{g}_{\mu\nu}$ as
\begin{eqnarray}
\hat{g}_{00}=-N^2+N_i g^{ij}N_j \ ,
\quad \hat{g}_{0i}=N_i \ , \quad
\hat{g}_{ij}=g_{ij} \ ,
\nonumber \\
\hat{g}^{00}=-\frac{1}{N^2} \ , \quad
\hat{g}^{0i}=\frac{N^i}{N^2} \ , \quad
\hat{g}^{ij}=g^{ij}-\frac{N^i N^j}{N^2}
\ .
\nonumber \\
\end{eqnarray}
Then it is easy to see that
\begin{equation}
\sqrt{-\det \hat{g}}=N\sqrt{\det g} \ ,
\quad
\hg^{\mu\nu}\partial_\mu\phi
\partial_\nu \phi=-(\nabla_n\phi)^2+g^{ij}\partial_i\phi
\partial_j\phi \ .
\end{equation}
Further, $D+1$-dimensional
curvature ${}^{(D+1)}R$   can be written as
\begin{equation}\label{RD1}
{}^{(D+1)}R=
K^{ij}K_{ij}-K^2+R^{(D)}+\frac{2}{\sqrt{-\hat{g}}}
\partial_\mu(\sqrt{-\hat{g}}n^\mu K)-
\frac{2}{\sqrt{g}N}\partial_i
(\sqrt{g}g^{ij}\partial_j N) \ ,
\end{equation}
where $K_{ij}=\frac{1}{2N}(\partial_t g_{ij}
-\nabla_i N_j-\nabla_j N_i)$ and  where
$\nabla_i$ is covariant derivative
defined using the metric $g_{ij}$.
Further, $K=g^{ij}K_{ji}$. In what
follows we ignore these boundary terms
when we will presume appropriate boundary
conditions.
Let us  consider the scalar field action.
Using the notation introduced above  we
find the momentum conjugate to $\phi$
and $\lambda$
\begin{equation}
p_\phi=\sqrt{g}(\omega+\lambda)\nabla_n\phi \ ,
\quad
p_\lambda\approx 0 \ .
\end{equation}
Then  the Hamiltonian for the scalar field takes
the form
\begin{eqnarray}\label{mHTphi}
H^\phi&=&\int d^D\bx \mH^\phi \ ,
\quad \mH^\phi
= N\mH_T^\phi+N^i\mH_i^\phi \ ,
\quad \mH_i=
p_\phi \partial_i \phi \ ,
\nonumber \\
\mH_T^\phi&=&\frac{1}{2\sqrt{g}(\omega+\lambda)}p_\phi^2+
\frac{1}{2}\sqrt{g}(\omega+\lambda)g^{ij}\partial_i\phi
\partial_j\phi+\sqrt{g}V+\sqrt{g}
\lambda U  \ .
\nonumber \\
\end{eqnarray}
Finally we write the
Hamiltonian for General Relativity part of the action
\begin{equation}
H^{GR}=\int d^D\bx \mH^{GR} \ , \quad
\mH^{GR}=N\mH_T^{GR}+N^i\mH_i^{GR}\  ,
\end{equation}
where
\begin{eqnarray}
\mH_T^{GR}&=&\frac{1}{\sqrt{g}}\pi^{ij}g_{ik}g_{jl}
\pi^{kl}-\frac{1}{\sqrt{g}D}\pi^2
-\sqrt{g}R^{(D)} \ ,
\nonumber \\
\mH_i^{GR}&=&-2 g_{ik}\nabla_j\pi^{kj} \ ,
\nonumber \\
\end{eqnarray}
where $\pi^{ij}$ is momentum conjugate to
$g_{ij}$ with non-trivial Poisson brackets
\begin{equation}
\pb{g_{ij}(\bx),\pi^{kl}(\by)}=
\frac{1}{2}(\delta_i^k\delta_j^l+\delta_i^l\delta_j^k)
\delta(\bx-\by) \ ,
\end{equation}
and where $\pi\equiv \pi^{ij}g_{ji}$. Note that
$\nabla_i$ is a covariant derivative calculated
with the metric $g_{ij}$ that also obeys $\nabla_i
g_{jk}=0$.

In summary, the total Hamiltonian is
$H=H^\phi+H^{GR}$. The preservation of the
primary constraints $p_N\approx 0 \ , p^i
\approx 0$ implies the secondary ones
\begin{equation}\label{MHTi}
\mH_T=\mH_T^{GR}+\mH_T^\phi\approx 0 \ ,
\quad \mH_i=\mH_i^{GR}+\mH_i^\phi \approx 0 \ .
\end{equation}
It is useful  to introduce the smeared form of these
constraints
\begin{eqnarray}
\bT_T(N)&=&\bT_T^{GR}(N)+\bT_T^{\phi}(N) \ , \nonumber \\
\bT_S(N^i)&=&\bT_S^{GR}(N^i)+\bT_S^{\phi}(N^i) \ , \nonumber \\
\end{eqnarray}
where
\begin{eqnarray}
\bT^{GR}_T(N)&=&\int d^D\bx N\mH_T^{GR} \  ,
\quad
\bT^{\phi}_T(N)=\int d^D\bx N\mH_T^{\phi} \  ,
\nonumber \\
\bT^{GR}_S(N^i)&=&\int d^D\bx N^i\mH^{GR}_i \ ,
\quad
\bT^{\phi}_S(N^i)=\int d^D\bx (N^i\mH^\phi_i+N^i
p_\lambda \partial_i\lambda) \ , \nonumber \\
\end{eqnarray}
where we included the primary constraint $p_\lambda\approx 0$
into definition of $\bT^\phi_S(N^i)$ in order to
ensure the correct form of the Poisson bracket
between the diffeomorphism generator $\bT_S^\phi(N^i)$
and the scalar field $\lambda$.

It is well known that  the Poisson brackets between
smeared form of the General Relativity constraints
take the form
\cite{Isham:1984sb,Isham:1984rz,Hojman:1976vp}
\begin{eqnarray}\label{pbconGR}
\pb{\bT^{GR}_T(N),\bT^{GR}_T(M)}&=&
\bT^{GR}_S(g^{ij}(N\partial_j M-M\partial_j N)) \ ,
\nonumber \\
\pb{\bT^{GR}_S(N^i),\bT^{GR}_T(M)}&=&
\bT^{GR}_T(N^i\partial_i M) \ , \nonumber \\
\pb{\bT^{GR}_S(N^i),\bT^{GR}_S(M^i)}&=&
\bT^{GR}_S(N^j\partial_j M^i-M^j\partial_j N^i) \ .
\nonumber \\
\end{eqnarray}
On the other hand we have to determine
the Poisson brackets between constraints
corresponding to the scalar field.
First of all it is easy to see that
\begin{equation}
\pb{\bT_S^\phi(N^i),
\bT_S^\phi(M^i)}=
\bT_S^\phi(N^j\partial_j M^i-
M^j\partial_j N^i) \ .
\end{equation}
On the other hand the Poisson
bracket between $\bT_S(N^i)$ and
$\bT_T^\phi(M)$ is equal to
\begin{eqnarray}\label{bTS}
\pb{\bT_S(N^i),\bT_T^\phi(M)}&=&
\int d^D\bx  (-N^k\partial_k\mH^\phi_T-
\partial_k N^k \mH_T^\phi)=\nonumber \\
&=&\int d^D \bx N^k\partial_k \mH_T^\phi=
\bT_T^\phi(N^k\partial_k M) \nonumber \\
\end{eqnarray}
using
\begin{eqnarray}
\pb{\bT_S(N^i),g_{ij}}&=&
-N^k\partial_k g_{ij}-\partial_i\xi^k g_{kj}-
g_{ik}\partial_j \xi^k \ , \nonumber \\
\pb{\bT_S(N^i),\sqrt{g}}&=&
-N^k\partial_k \sqrt{g}-\sqrt{g}\partial_k N^k
\ . \nonumber \\
\end{eqnarray}
Note that the presence of the
term  $N^i p_\lambda\partial_i\lambda$ in
the definition of $\bT_S^\phi(N^i)$
was crucial for  deriving of  the correct
form of the Poisson bracket (\ref{bTS}).
Finally we calculate the Poisson bracket
between $\bT_T^\phi(N),\bT_T^\phi(M)$ and
after some algebra we find the desired
result
\begin{equation}
\pb{\bT_T^\phi(N),\bT_T^\phi(M)}=
\bT_S^\phi(g^{ij}(N\partial_j M-
M\partial_j N)) \ .
\end{equation}
It is also easy to show that
\begin{eqnarray}
\pb{\bT_T^{GR}(N),\bT_T^\phi(M)}+
\pb{\bT_T^\phi(N),\bT_T^{GR}(M)}=0
\nonumber \\
\end{eqnarray}
due to the fact that $\mH_T^\phi$
depends on $g$ and not on their derivatives.
If we combine these results we find
that the Poisson brackets of the
constraints $\bT_T(N),\bT_S(N^i)$
has the desired
form (\ref{pbconGR}).

As the next step we analyze the stability
of   the primary constraint
  $p_\lambda\approx 0$. The requirement of its
  stability implies  the
 secondary constraint
\begin{eqnarray}\label{defG}
\partial_t p_\lambda(\bx)=
\pb{p_\lambda(\bx),H}=
\frac{1}{2\sqrt{g}(\omega+\lambda)^2}p_\phi^2-
\frac{1}{2}\sqrt{g}g^{ij}\partial_i\phi
\partial_j\phi-\sqrt{g}U\equiv \mG_\lambda(\bx)\approx 0 \ .
\nonumber \\
\end{eqnarray}
We observe that
\begin{equation}
\pb{p_\lambda(\bx),\mG_\lambda(\by)}=
\frac{1}{\sqrt{g}(\omega+\lambda)^3}p_\phi^2(\bx)
\delta(\bx-\by) \ .
\end{equation}
In other words  $p_\lambda$ and $\mG_\lambda$ are
the second class constraints.
However there are additional  non-zero Poisson brackets.
The first one is
\begin{eqnarray}
\pb{\mG_\lambda(\bx),\mG_\lambda(\by)}&=&
-2\frac{1}{\sqrt{g}(\omega+\lambda)^2}
g^{ij}\mH_j^\phi(\bx)\partial_i\delta(\bx-\by)-\nonumber \\
&-&\partial_i \left[\frac{1}{\sqrt{g}(\omega+\lambda)^2}
g^{ij}\mH_j^\phi(\bx)\right]\delta(\bx-\by)  \  .
\nonumber \\
\end{eqnarray}
It is also clear from the structure of the
constraint $\mG_\lambda$ that there is
 non-zero Poisson
brackets between $\mG_\lambda$ and $\mH$
defined by (\ref{MHTi})
\begin{eqnarray}
\pb{\mG_\lambda(\bx),\mH}\neq 0 \ ,
\end{eqnarray}
where $\mH=N\mH_T+N^i\mH_i$. Note that
the explicit form of this Poisson bracket
is not important for us.

Using these results we can proceed to the
study of the stability of the secondary
constraints.
Following the  standard analysis of the constraint
systems we introduce the total Hamiltonian
as
\begin{equation}
H_T=H+\int d^D\bx (\alpha
 p_\lambda+\beta \mG_\lambda) \ ,
 \end{equation}
 where $\alpha,\beta$ are Lagrange multipliers
 and analyze the stability of the constraints
$\mH, \mG_\lambda,p_\lambda$.
 Firstly we have
 \begin{eqnarray}
 \partial_t \mH&=&
 \pb{\mH(\bx),H}+
 \int d^D\by (\beta\pb{\mH(\bx),\mG_\lambda(\by)}
+\beta \pb{\mH(\bx),p_\lambda(\by)})\approx
\nonumber \\
& &\approx \int d^D \by \beta
\pb{\mH(\bx),\mG_\lambda(\by)}\neq 0 \ ,
\end{eqnarray}
where we used the fact that the Poisson
brackets between $\mH$ and $H$ weakly
vanish. Then the requirement of stability of the constraint
$\mH\approx 0$ determines the value of the
 Lagrange
multiplier $\beta$ to be equal to  $0$. On the other
hand the time evolution of the constraint
$\mG_\lambda$ is given by the equation
\begin{eqnarray}
\partial_t \mG_\lambda(\bx)=
\pb{\mG_\lambda(\bx),H}+\int d^D\by\alpha(\by)
\pb{\mG_\lambda(\bx),p_\lambda(\by)}\approx 0 \ .
\nonumber \\
\end{eqnarray}
Due to the fact that $\pb{\mG_A,H}\neq 0$ and
$\pb{\mG_\lambda,p_\lambda}\neq 0$ the equation
above can be solved for  $\alpha$ at least
in principle.  Then using these results
it is easy to see that the constraint
 $p_\lambda\approx 0$ is preserved during
 the time evolution of the system.
Further, $p_\lambda$ and $\mG_\lambda$ are
the second class constraints that can be solved
for $\lambda$ and $p_\lambda$ so that the
reduced phase space is spanned by
$(g_{ij},\pi^{ij}),(\phi,p_\phi)$ and
the symplectic structure is given
by the Dirac brackets
between these variables. In order to
find their form we
 introduce following notation for the
Poisson brackets of the second class constraints
$p_\lambda,\mG_\lambda$
\begin{eqnarray}\label{triangle}
\triangle_{11}(\bx,\by)&=&
\pb{p_\lambda(\bx),p_\lambda(\by)}=0 \ , \quad
\triangle_{12}(\bx,\by)=
\pb{p_\lambda(\bx),\mG_\lambda(\by)}\neq 0 \ , \nonumber \\
\triangle_{21}(\bx,\by)&=&
\pb{\mG_\lambda(\bx),p_\lambda(\by)}\neq 0 \ , \quad
\triangle_{22}(\bx,\by)=\pb{\mG_\lambda(\bx),\mG_\lambda(\by)}\neq 0 \
\nonumber \\
\end{eqnarray}
and  denote the inverse matrix as $(\triangle^{-1})^{AB}(\bx,\by)$.
This matrix  by definition obeys the equation
\begin{equation}
\int d^\bx \triangle_{AC}(\bx,\bz)
(\triangle^{-1})^{CB}(\bz,\by)=
\delta_A^B\delta(\bx-\by) \ .
\end{equation}
It can be shown that the matrix $(\triangle^{-1})$
has following structure
\begin{equation}
(\triangle^{-1})=\left(\begin{array}{cc}
* & * \\
* & 0 \\ \end{array}\right) \ ,
\end{equation}
where $*$ denotes non-zero elements.
It is important for the calculation of the
Dirac brackets
that $(\triangle^{-1})^{22}=0$. Explicitly, the
Dirac bracket between $\phi$ and $p_\phi$ takes the form
\begin{eqnarray}
\pb{\phi(\bx),p_\phi(\by)}_D &=&
\pb{\phi(\bx),p_{\phi}(\by)}-\nonumber \\
&-&\int d^D\bz d^D \bz'
\pb{\phi(\bx),\Phi_A(\bz)}(\triangle^{-1})^{AB}
(\bz,\bz') \pb{\Phi_B(\bz'), p_\phi(\by)}=
\nonumber \\
&=&\pb{\phi(\bx),p_{\phi}(\by)}-\int d^D\bz d^D \bz'
\pb{\phi(\bx),\mG_\lambda(\bz)}(\triangle^{-1})^{22}
(\bz,\bz') \pb{\mG_\lambda(\bz'), p_\phi(\by)}=
\nonumber \\
&=&\pb{\phi(\bx),p_{\phi}(\by)} \ ,  \nonumber \\
\end{eqnarray}
where $\Phi_A=(p_\lambda,\mG_\lambda)$ is the
common notation for the second class constraints.

We are now ready to completely eliminate the
second class constraints $\Phi_A$.  The
 constraint $\mG_\lambda=0$ can be solved for
 $\omega+\lambda$
\begin{equation}
(\omega+\lambda)=
\frac{p_\phi}{\sqrt{g}\sqrt{g^{ij}\partial_i\phi
\partial_j\phi+2 U}} \ .
\end{equation}
Inserting this result into the Hamiltonian
constraint (\ref{mHTphi}) we find that
it takes the form
\begin{equation}\label{mHphired}
\mH_T^\phi=
p_\phi\sqrt{g^{ij}\partial_i\phi\partial_j\phi+
2U}+\sqrt{g}V-\sqrt{g}\omega U \ .
\end{equation}
We observe that this Hamiltonian density  is linear
in momenta. Then  the equation of motion
for $\phi$ takes the form
\begin{equation}
\partial_t\phi=\pb{\phi,H}=
N\sqrt{g^{ij}\partial_i\phi\partial_j\phi+
2U}+N^i\partial_i\phi
\end{equation}
that shows that the time evolution  of $\phi$
does not depend on $p_\phi$. Such systems
were extensively studied in the past in the
context of 't Hooft's deterministic approach
to quantum mechanics
\cite{'tHooft:2001ct,'tHooft:1999gk,'tHooft:1999bx,Blasone:2009yp}
and it is really interesting that the
Hamiltonian with similar structure arises
in Lagrange modified multiplier theory.
\footnote{We review the basic
facts considering such system, following
\cite{Blasone:2004yf}.
Let us consider the Hamiltonian system
\begin{equation}\label{Hq}
H=p_i f^i(q)+U(q)  \ , i=1,\dots,N \ .
\end{equation}
From (\ref{Hq}) we determine
 the equations of motion for $q^i$
 \begin{equation}\label{eqf}
\partial_t q^i=\pb{q^i,H}=f^i(q) \ .
\end{equation}
This equation  for $q^i$
is autonomous, i.e., it is decoupled from the conjugate momenta
$p_i$.
 Further it is impossible to perform the
Legendre transformation to the Lagrangian
since $H_{ij}=
\frac{\partial^2 H}{\partial p_i\partial p_j}=0$.
However it is possible to find
 the Lagrangian that gives the
equation of motion (\ref{eqf}) when we introduce
the auxiliary fields $\lambda_i$ and write
the Lagrangian as
\begin{equation}\label{Llambda}
L=\lambda_i (\dot{q}^i-f^i(q))-U(q) \ .
\end{equation}
No we show that from  (\ref{Llambda})
we can derive the Hamiltonian
 (\ref{Hq}). The momenta conjugate
to $\lambda_i$ and $q^i$ take the form
\begin{equation}
p^i_\lambda=\frac{\delta L}{\delta
\dot{\lambda}_i}\approx 0 \ ,
\quad
p_i^q=\frac{\delta L}{\delta \dot{q}^i}=\lambda_i
\end{equation}
so that we have two sets of primary
constraints
\begin{equation}
\Phi^i_\lambda=p^i_\lambda\approx 0 \ ,
\quad
\Phi_i^q=p^q_i-\lambda_i\approx 0 \ .
\end{equation}
The extended Hamiltonian  that follows
from (\ref{Llambda})
takes the form
\begin{equation}
H_E=H+\omega_i^\lambda \Phi^i_\lambda+
\omega^i_q\Phi^q_i \ , \quad  H=\lambda_i f^i+U(q) \ .
\end{equation}
Then we  study the stability of the constraints
$\Phi_\lambda^i \ , \Phi_i^q$
\begin{eqnarray}
\partial_t \Phi^i_\lambda&=&
\pb{\Phi^i_\lambda,H_E}=-f^i+\omega^i_q=
0 \nonumber \\
\partial_t \Phi_i^q&=&
\pb{\Phi_i^q,H_E}=-\lambda_j
\frac{d f^j}{dq^i}-\omega_i^q=0 \ .
\nonumber \\
\end{eqnarray}
From these equations we
can in principle  determine the
Lagrange multipliers. In other words the
constraints $\Phi_\lambda^i ,\Phi_i^q$
are the second class that should strongly
vanish. The solving of these constraints
we find the Hamiltonian
\begin{equation}
H=p_i^q f^i+U(q) \
\end{equation}
that coincides with the Hamiltonian
(\ref{Hq}). Further, it can be easily shown
that the Dirac brackets between
$q^i$ and $p_i$ coincide with their
Poisson brackets. However the problem
with the Hamiltonian (\ref{Hq}) is
that is not bounded from below which is
due to the absence of a leading kinetic term quadratic in the
momenta $(p_i)^2$.}
We  complete our analysis by calculation of
the Poisson bracket between $\mH_T^\phi$
given in (\ref{mHphired}) and the
spatial diffeomorphism constraint $\bT_S(N^i)$.
Using
\begin{equation}
\pb{\bT_S(N^i),g^{ij}\partial_i\phi
\partial_j\phi}=
-N^k\partial_k \left(g^{ij}\partial_i\phi
\partial_j\phi\right)
\end{equation}
we easily find
\begin{equation}
\pb{\bT_S(N^i),\mH_T^\phi}=
-N^i\partial_i \mH_T^\phi-
\partial_i N^i \mH_T^\phi \ ,
\end{equation}
where $\mH_T^\phi$ was given in (\ref{mHphired}).
The analysis of the remaining Poisson brackets
is the same as above with conclusion that
the smeared form of the constraints
obey the algebra of constraints  given in
(\ref{pbconGR}). In other words we show that
the Lagrangian multiplier modified scalar action
together with General Relativity action
obeys the basis rules of geometrodynamics.
%
\section{Hamilton Analysis of $F(R)$
Theories with Lagrange Multipliers}
\label{third}
It turns out that the Lagrange multiplier
modified $F(R)$-gravity possesses many interesting
properties. For example, the reconstruction
programme can be more easily performed in Lagrange
multiplier modified gravity
\cite{Capozziello:2010uv}.  In usual
$F(R)$-gravity, we need to
 solve the complicated differential
  equation     to realize the reconstruction program,
   for recent review, see \cite{Nojiri:2010ny}.
It was demonstrated in \cite{Capozziello:2010uv}
 that the presence of constraint significantly
  simplifies the reconstruction scenario.
It was also shown there that the  presence of
 Lagrange multiplier  implies that it is
 necessary to include the second $F(R)$ function into
 action.

The action introduced in
\cite{Capozziello:2010uv} takes the form
\begin{equation}\label{SFF}
S=\int d^{D+1}x
\sqrt{-\hg}\left[F_1({}^{(D+1)}R)-
\lambda \left(\frac{1}{2}\partial_\mu {}^{(D+1)}R
\hg^{\mu\nu}\partial_\nu {}^{(D+1)} R+F_2({}^{(D+1)}R)\right)\right] \ .
\end{equation}
Introducing two auxiliary fields $A,B$ we
can rewrite the action  (\ref{SFF}) into the
form
\begin{equation}\label{SFF2}
S=\int d^{D+1}x
\sqrt{-\hg}\left[F_1(A)-
\lambda \left(\frac{1}{2}\partial_\mu A
\hg^{\mu\nu}\partial_\nu A+F_2(A)\right)+
B({}^{(D+1)}R-A)\right]  \ .
\end{equation}
It is easy to see that integration of
$A,B$ from (\ref{SFF2}) leads to (\ref{SFF}).
Our goal is to find the Hamiltonian from
(\ref{SFF2}) implementing
$D+1$ formalism.
In fact using (\ref{RD1})
it is easy to see that the action
(\ref{SFF2}) takes
the form
\begin{eqnarray}\label{Sextend}
S&=&\int d^D\bx dt
\sqrt{g}N\left( F_1(A)-
\lambda (-\nabla_n A\nabla_n A+
g^{ij}\partial_i A\partial_j A
+F_2(A))-
BA\right)+ \nonumber \\
&+&
\int d^D\bx dt \sqrt{g}NB(K_{ij}\mG^{ijkl}K_{kl}
+R^{(D)}-A)-\nonumber \\
&-&
-2\int d^D\bx dt(\sqrt{g}(\partial_t B-N^i\partial_i B) K
+2\sqrt{g}\partial_i B g^{ij}\partial_j N ) \ ,
\nonumber \\
\end{eqnarray}
where we performed integration by parts and
ignored boundary terms.
From (\ref{Sextend}) we easily find momenta conjugate
to canonical variables $g_{ij},N,N_i,A$ and $B$
\begin{eqnarray}
\pi^{ij}&=&
\sqrt{g}B\mG^{ijkl}K_{kl}-\sqrt{g}
\nabla_n Bg^{ij} \ , \quad, p_N\approx 0 \ ,
\quad ,p^i \approx 0 \ ,
\nonumber \\
p_B&=&-2\sqrt{g}K \ , \quad  p_A=2\sqrt{\lambda}
\nabla_n A \ , \quad p_\lambda\approx 0  \ . \nonumber \\
\end{eqnarray}
Note that the  Lagrange
multiplier  implies that  $A$ is a dynamical field
which is different from standard $F(R)$ theory of
gravity where $A$ remains auxiliary field.
Then  after some effort we derive
the  Hamiltonian density in the form
\begin{eqnarray}
\mH&=& N\mH_T+N^i\mH_i \ ,
\nonumber \\
\end{eqnarray}
where
\begin{eqnarray}\label{mHTA}
\mH_T&=&\frac{1}{\sqrt{g}B}\pi^{ij}g_{ik}g_{il}\pi^{kl}
-\frac{1}{\sqrt{g}BD}\pi^2-\frac{\pi p_B}{\sqrt{g}D}
+\nonumber \\
&+&\frac{B}{4\sqrt{g}D}(D-1)p_B^2-\sqrt{g}BR^{(D)}+
2\partial_i[\sqrt{g}g^{ij}\partial_j B]
\nonumber \\
&+&\frac{1}{4\sqrt{g}\lambda}p_A^2
+\sqrt{g}BA -
\sqrt{g}[F_1(A)-
\lambda (g^{ij}\partial_i A\partial_j A
+F_2(A))] \ ,  \nonumber \\
\end{eqnarray}
and where
\begin{equation}
\mH_i=p_A\partial_i A+p_B\partial_iB+p_\lambda
\partial_i\lambda
-2g_{ik}\nabla_j \pi^{jk} \ .
\end{equation}
For further purposes we split  $\mH_T$ into two parts
as $\mH_T=\mH_T^{GR}+
\mH_T^{A}$ where
\begin{eqnarray}\label{mHT3}
\mH_T^{GR}&=&
\frac{1}{\sqrt{g}B}\pi^{ij}g_{ik}g_{il}\pi^{kl}
-\frac{1}{\sqrt{g}BD}\pi^2-\frac{\pi p_B}{\sqrt{g}D}
+\nonumber \\
&+&\frac{B}{4\sqrt{g}D}(D-1)p_B^2-\sqrt{g}BR^{(D)}+
2\partial_i[\sqrt{g}g^{ij}\partial_j B]
\nonumber \\
\mH_T^A&=&\frac{1}{4\sqrt{g}\lambda}p_A^2
+\sqrt{g}BA -
\sqrt{g}[F_1(A)-
\lambda (g^{ij}\partial_i A\partial_j A
+F_2(A))] \  .  \nonumber \\
\end{eqnarray}
 The theory possesses $2+D$
primary constraints
\begin{equation}
\pi_N\approx 0 \ , \pi_i\approx 0 \ ,
\pi_\lambda \approx 0 \ .
\end{equation}
The preservation of the primary constraints
$\pi_N$ and $\pi_i$ imply the secondary
constraints $\mH_T\approx 0 \ , \mH_i\approx 0
$ while the preservation of $\pi_\lambda\approx 0$
leads to the secondary constraint
\begin{equation}\label{mGlambda2}
\mG_\lambda=\frac{1}{4\sqrt{g}\lambda^2}
p_A^2-\sqrt{g}(g^{ij}\partial_i A\partial_j A+F_2(A))
\approx 0
\end{equation}
We see that it takes the same form as the secondary
constraint (\ref{defG}). Clearly $p_\lambda$ together
with $\mG_\lambda$ are the  second class constraints.
Properties of these constraints were analyzed in
previous section and results derived there can
be used in this section as well.

On the other hand the form of the Hamiltonian
constraint $\mH_T^{GR}$ is new and we have
to check that this constraint is
preserved during the time evolution of
the system. In other words we have to
calculate the Poisson brackets of the
smeared form of these constraints
\footnote{In
\cite{Chaichian:2010zn} similar analysis has been
performed in the context of non-projectable
version of Ho\v{r}ava-Lifshitz $F(R)$ gravity.}
\begin{equation}
\bT_T^{GR}(N)=\int d^D\bx N(\bx) \mH_T^{GR}(\bx) \ ,
\quad \bT_S^{GR}(N^i)
=\int d^D\bx N^i(\bx) \mH_i^{GR}(\bx) \ .
\end{equation}
Let us now outline the strategy of the
calculations of these Poisson brackets.
In the process  of their calculations
several delta functions occur.  However
it turns out that the non-zero contributions
give terms  that contain
derivatives of these delta functions. Such
expressions arise for example from following
Poisson bracket
\begin{equation}
\pb{\pi^{kl}(\bx), (\sqrt{g}R^D)(\by)}=
-\frac{\delta (\sqrt{g}R^{(D)}(\by))}{\delta
g_{kl}(\bx)}  \ .
\end{equation}
The right side of this equation
can be calculated using the formulas
\begin{eqnarray}
\delta R^{(D)}=
-(R^{(D)})^{ij}\delta g_{ij}
+\nabla^i\nabla^j \delta g_{ij}
-g^{ij}\nabla_k\nabla^k\delta g_{ji} \ ,
\quad
\delta g=g g^{ij}\delta g_{ij} \ .
\nonumber \\
\end{eqnarray}
Now we are ready to perform these
calculations. It turns out that
following non-zero Poisson
brackets contribute to the final
result
\begin{eqnarray}
& & - \pb{\int d^D\bx N \frac{1}{B\sqrt{g}}
\pi^{ij}g_{ik}g_{jl}\pi^{ij},
\int d^D\by M \sqrt{g}R^{(D)}}-
\nonumber \\
&-&\pb{\int d^D\by N \sqrt{g}R^{(D)},
\int d^D\bx M \frac{1}{B\sqrt{g}}
\pi^{ij}g_{ik}g_{jl}\pi^{ij}}=
\nonumber \\
&=&2\int d^D\bx (N\nabla_i \nabla_j M-M\nabla_i \nabla_j N)
\pi^{ij}
-2\int d^D\bx \pi (N\nabla_i \nabla^i M-M\nabla_i \nabla^i N)+
\nonumber \\
&+&4\int d^D \bx \pi^{ij}(N\nabla_i M-M\nabla_i N)\pi^{ij}\frac{1}{B}
\nabla_j B-4\int d^D\bx \pi (N\nabla_i M-M\nabla_i N)\frac{1}{B}
\nabla^i B \ ,
\nonumber \\
\end{eqnarray}
\begin{eqnarray}
& &\pb{\int d^D\bx
\frac{N}{\sqrt{g}B}\pi^2,
\int d^D\by M\sqrt{g}R^{(D)}}+
\pb{\int d^D\bx N\sqrt{g}R^{(D)},
\int d^D\by
\frac{M}{\sqrt{g}B}\pi^2
}=
\nonumber \\
&=&-\frac{2}{D}\int d^D\bx
\pi (N\nabla_i \nabla^i M-M\nabla_i
\nabla^i N)+
2\int d^D\bx
\pi (N\nabla_i \nabla^i M-M\nabla_i
\nabla^i N)-\nonumber \\
&-&\frac{4}{D}\int d^D\bx \pi (N\nabla_i M-M\nabla_i N)
\frac{\nabla^i B}{B}
+4\int d^D\bx \pi (N\nabla_i M-M\nabla_i N)
\frac{\nabla^i B}{B}
\nonumber \\
\end{eqnarray}
and
\begin{eqnarray}
& &\pb{\int d^D\bx N\frac{\pi p_B}{D\sqrt{g}},
\int d^D\by \sqrt{g}R^{(D)} BM}
+\pb{\int d^D\by \sqrt{g}R^{(D)} BN,
\int d^D\bx M\frac{\pi p_B}{D\sqrt{g}}}
=\nonumber \\
&=&-\frac{(1-D)}{D}\int d^D\bx
p_B B( N\nabla_i \nabla^i M-M\nabla_i\nabla^i N)
-\nonumber \\
&-&\frac{2(1-D)}{D}
\int d^D\bx
p_B ( N\nabla_i M M-M\nabla_i N)
\nabla^i B \ ,
\nonumber \\
\end{eqnarray}
\begin{eqnarray}
& &\pb{\int d^D\bx
\frac{N}{B\sqrt{g}}
\pi^{ij}g_{ik}g_{jl}\pi^{kl},\int d\by
2M\partial_i [\sqrt{g}g^{ij}\partial_j B]}+
\nonumber \\
&+& \pb{
\int d\by
2N\partial_i [\sqrt{g}g^{ij}\partial_j B],
\int d^D\bx
\frac{M}{B\sqrt{g}}
\pi^{ij}g_{ik}g_{jl}\pi^{kl}}=
\nonumber \\
&=&2\int d^D\bx \pi \frac{1}{B} (N\nabla_i
M-M\nabla_i N)g^{ij}\nabla_j B-
4\int d^D\bx \frac{1}{B}(N\nabla_i M-M\nabla_i N)
\pi^{ij}\nabla_j B \nonumber \\
\end{eqnarray}
and
\begin{eqnarray}
&- &\pb{
\int d^D\bx \frac{N}{\sqrt{g}BD}
\pi^2, \int d^D\by M2\partial_m[\sqrt{g}
g^{mn}\partial_n B]}-\nonumber \\
&-&
\pb{ \int d^D\bx N2\partial_m[\sqrt{g}
g^{mn}\partial_n B],
\int d^D\bx \frac{M}{\sqrt{g}BD}
\pi^2}=\nonumber \\
&=&\frac{2(2-D)}{D}
\int d^D\bx \frac{\pi}{B}(N\nabla_i
M-M\nabla_i N)\nabla^i B
\nonumber \\
\end{eqnarray}
and
\begin{eqnarray}
&-&\pb{\int d^D\bx
\frac{N}{\sqrt{g}D}\pi p_B,
\int d^D\by 2M\partial_m[\sqrt{g}g^{mn}
\partial_n B]}+\nonumber \\
&-&\pb{\int d^D\by 2 N\partial_m[\sqrt{g}g^{mn}
\partial_n B],
\int d^D\bx
\frac{M}{\sqrt{g}D}\pi p_B}=
\nonumber \\
&=&\frac{(2-D)}{D}\int d^D\bx
(N\nabla_m M-M\nabla_m N)p_B\nabla^m B
+\nonumber \\
&+&\frac{2}{D}\int d^D\bx
\pi ( N \nabla_m \nabla^m M-M\nabla_m \nabla^m N)  \ ,
\nonumber \\
\end{eqnarray}
\begin{eqnarray}
& &\pb{\int d^D\bx
\frac{NB}{4\sqrt{g}D}(D-1)p_B^2,
2\int d^D\by M \partial_m[\sqrt{g}g^{mn}\partial_n B]}+
\nonumber \\
&+&\pb{
2\int d^D\bx N\partial_m[\sqrt{g}g^{mn}\partial_n B]
,\int d^D\by
\frac{MB}{4\sqrt{g}D}(D-1)p_B^2}=
\nonumber \\
&=&-\frac{D-1}{D}
\int d^D\bx p_B B(N\nabla_m \nabla^m M-
M\nabla_m \nabla^m N) \ .  \nonumber \\
\end{eqnarray}
Collecting all these terms together we obtain
that almost all contributions  cancel and the final result
takes the form
\begin{equation}
\pb{\bT^{GR}_T(M),\bT^{GR}_T(N)}=
\bT^{GR}_S((N\nabla_j M-M\nabla_j N)g^{ji}) \ .
\end{equation}
In other words the Poisson bracket
of the smeared form of the Hamiltonian constraints
(\ref{mHT3}) has the same form as in General
Relativity and hence it is  with agreement
with basic principles of geometrodynamics. Alternatively,
it has the form that is expected for fully diffeomorphism
invariant theory. Note also that the Poisson
bracket between smeared form of the diffeomorphism
and Hamiltonian constraint takes the standard form
that follows from the fact that Hamiltonian is
manifestly invariant under spatial diffeomorphism.
Then it is clear that the diffeomorphism and Hamiltonian
constraints  are preserved during the time evolution
of the system.

Now it is straightforward to finish the
analysis of the Poisson brackets of the
constraints of the Lagrange multiplier
modified gravity. Since the
Poisson brackets of the
constrains corresponding to the gravity
part of the action   are the same as in General
Relativity and since the scalar part of the
constraints has exactly the same form
as in previous section
 we immediately find that
the Poisson brackets of the Lagrange multiplier
modified $F(R)$ gravity take the form
\begin{eqnarray}\label{pbfull}
\pb{\bT_T(N),\bT_T(M)}&=&
\bT_S(g^{ij}(N\partial_j M-M\partial_j N)) \ ,
\nonumber \\
\pb{\bT_S(N^i),\bT_T(M)}&=&
\bT_T(N^i\partial_i M) \ , \nonumber \\
\pb{\bT_S(N^i),\bT_S(M^i)}&=&
\bT_S(N^j\partial_j M^i-M^j\partial_j N^i) \ .
\nonumber \\
\end{eqnarray}
where
\begin{equation}
\mH_T=\mH_T^{GR}+\mH_T^A \ ,  \quad
\mH_i=-g_{il}\nabla_k \pi^{lk}+p_A\partial_i A \ ,
\end{equation}
where $\mH_T^{GR} $ is given in
(\ref{mHTA}). Note that  $\mH_T^A$
is equal to
\begin{equation}
\mH_T^A=p_A
\sqrt{g^{ij}\partial_i A\partial_j A+F_2(A)}+
\sqrt{g}BA-\sqrt{g}F_1(A) \
\end{equation}
after solving the second class constraint
$\mG_\lambda $ given in
(\ref{mGlambda2})
with respect to $\lambda$
\begin{equation}
\lambda=\frac{p_A}{2\sqrt{g}
\sqrt{F_2(A)+g^{ij}\partial_i \phi\partial_j\phi}} \ .
\end{equation}
 \noindent {\bf
Acknowledgements:}

 This work   was also
supported by the Czech Ministry of
Education under Contract No. MSM
0021622409.
 I would like also
thank to Max Planck Institute at Golm
for its financial support and
 kind hospitality during my work on
this project.

\vskip 5mm



\begin{thebibliography}{20}



\bibitem{Gao:2010gj}
  C.~Gao, Y.~Gong, X.~Wang and X.~Chen,
\emph{``Cosmological
models with Lagrange Multiplier Field,''}
  arXiv:1003.6056 [astro-ph.CO].

\bibitem{Lim:2010yk}
  E.~A.~Lim, I.~Sawicki and A.~Vikman,
\emph{``Dust of Dark Energy,''}
  JCAP {\bf 1005} (2010) 012
  [arXiv:1003.5751 [astro-ph.CO]].

\bibitem{Capozziello:2010uv}
  S.~Capozziello, J.~Matsumoto, S.~Nojiri and S.~D.~Odintsov,
\emph{``Dark energy
from modified gravity with Lagrange multipliers,''}
  arXiv:1004.3691 [hep-th].



\bibitem{Du:2010rv}
  Y.~Du, H.~Zhang and X.~Z.~Li,
\emph{``A new mechanism to cross the phantom divide,''}
  arXiv:1008.4421 [astro-ph.CO].

\bibitem{Feng:2010tf}
  C.~J.~Feng and X.~Z.~Li,
\emph{``Non-Gaussianity with
 Lagrange Multiplier Field in the Curvaton Scenario,''}
  arXiv:1008.1152 [astro-ph.CO].


\bibitem{Cai:2010zm}
  Y.~F.~Cai and E.~N.~Saridakis,
\emph{``Cyclic cosmology
 from Lagrange-multiplier modified gravity,''}
  arXiv:1007.3204 [astro-ph.CO].




\bibitem{Nojiri:2010kx}
  S.~Nojiri and S.~D.~Odintsov,
\emph{``Covariant power-counting renormalizable gravity: Lorentz symmetry breaking
  and accelerating early-time FRW universe,''}
  arXiv:1007.4856 [hep-th].


\bibitem{Nojiri:2010tv}
  S.~Nojiri and S.~D.~Odintsov,
\emph{``A proposal for
 covariant renormalizable field theory of gravity,''}
  Phys.\ Lett.\  B {\bf 691} (2010) 60
  [arXiv:1004.3613 [hep-th]].


\bibitem{Nojiri:2009th}
  S.~Nojiri and S.~D.~Odintsov,
\emph{``Covariant
 Horava-like renormalizable gravity and its FRW cosmology,''}
  Phys.\ Rev.\  D {\bf 81} (2010) 043001
  [arXiv:0905.4213 [hep-th]].





\bibitem{Isham:1984sb}
  C.~J.~Isham and K.~V.~Kuchar,
\emph{``Representations Of Space-Time Diffeomorphisms. 1. Canonical Parametrized
  Field Theories,''}
  Annals Phys.\  {\bf 164} (1985) 288.

\bibitem{Isham:1984rz}
  C.~J.~Isham and K.~V.~Kuchar,
\emph{``Representations
 Of Space-Time Diffeomorphisms. 2. Canonical
Geometrodynamics,''}
  Annals Phys.\  {\bf 164} (1985) 316.








\bibitem{Hojman:1976vp}
  S.~A.~Hojman, K.~Kuchar and C.~Teitelboim,
\emph{``Geometrodynamics Regained,''}
  Annals Phys.\  {\bf 96} (1976) 88.


\bibitem{'tHooft:2001ct}
  G.~'t Hooft,
\emph{``Quantum mechanics and determinism,''}
  arXiv:hep-th/0105105.

\bibitem{'tHooft:1999gk}
  G.~'t Hooft,
\emph{``Quantum gravity
 as a dissipative deterministic system,''}
  Class.\ Quant.\ Grav.\  {\bf 16} (1999) 3263
  [arXiv:gr-qc/9903084].

\bibitem{'tHooft:1999bx}
  G.~'t Hooft,
\emph{``Determinism and
 dissipation in quantum gravity,''}
  arXiv:hep-th/0003005.


\bibitem{Blasone:2009yp}
  M.~Blasone, P.~Jizba and F.~Scardigli,
\emph{``Can quantum
 mechanics be an emergent phenomenon?,''}
  J.\ Phys.\ Conf.\ Ser.\  {\bf 174} (2009) 012034
  [arXiv:0901.3907 [quant-ph]].


\bibitem{DeFelice:2010aj}
  A.~De Felice and S.~Tsujikawa,
\emph{``f(R) theories,''}
  Living Rev.\ Rel.\  {\bf 13} (2010) 3
  [arXiv:1002.4928 [gr-qc]].


\bibitem{Blasone:2004yf}
  M.~Blasone, P.~Jizba and H.~Kleinert,
\emph{``Path Integral
 Approach to 't Hooft's Derivation of Quantum from Classical
Physics,''}
  Phys.\ Rev.\  A {\bf 71} (2005) 052507
  [arXiv:quant-ph/0409021].




\bibitem{Faraoni:2008mf}
  V.~Faraoni,
\emph{``f(R) gravity: successes and
challenges,''}
  arXiv:0810.2602 [gr-qc].


\bibitem{Nojiri:2006ri}
  S.~Nojiri and S.~D.~Odintsov,
\emph{``Introduction to modified
gravity and gravitational alternative
for dark energy,''}
  eConf {\bf C0602061} (2006) 06
  [Int.\ J.\ Geom.\ Meth.\ Mod.\ Phys.\  {\bf 4} (2007) 115]
  [arXiv:hep-th/0601213].


\bibitem{Deruelle:2009pu}
  N.~Deruelle, Y.~Sendouda and A.~Youssef,
\emph{``Various Hamiltonian
 formulations of f(R) gravity and their canonical
relationships,''}
  Phys.\ Rev.\  D {\bf 80} (2009) 084032
  [arXiv:0906.4983 [gr-qc]].

\bibitem{Chaichian:2010zn}
  M.~Chaichian, M.~Oksanen and A.~Tureanu,
\emph{``Hamiltonian analysis of non-projectable
modified F(R) Ho\v{r}ava-Lifshitz
  gravity,''}
  arXiv:1006.3235 [hep-th].

\bibitem{Nojiri:2010ny}
  S.~Nojiri and S.~D.~Odintsov,
\emph{``Non-singular modified
 gravity unifying inflation with late-time
acceleration and universality of viscous ratio bound in F(R) theory,''}
  arXiv:1008.4275 [hep-th].

\bibitem{Ezawa:1998ax}
  Y.~Ezawa, M.~Kajihara, M.~Kiminami, J.~Soda and T.~Yano,
\emph{``On the canonical formalism for
a higher-curvature gravity,''}
  Class.\ Quant.\ Grav.\  {\bf 16}, 1127 (1999)
  [arXiv:gr-qc/9801084].

\bibitem{Ezawa:1999ty}
  Y.~Ezawa, M.~Kajihara, M.~Kiminami, T.~Yano and J.~Soda,
\emph{``Semiclassical approach to
stability of the extra-dimensional
spaces in higher-curvature gravity
theories,''}
  Class.\ Quant.\ Grav.\  {\bf 16}, 1873 (1999).



\end{thebibliography}
\end{document}